\documentclass[12pt]{article}
\begin{document}
\title{OPTICAL KERR TYPE NONLINEARITY}
\author{Jaros\a{l}aw Zale\'sny \footnote{e-mail: jarek@ps.pl} \\  \\
\emph{Institute of Physics, Technical University of Szczecin,}   \\
\emph{Al. Piast\'ow 48, 70-310 Szczecin, Poland}}
\date{\today}
\maketitle

\begin{abstract}
The origin of Kerr type nonlinearity of the medium as a result of the interaction between photons via Dirac delta-potential is presented in the formalism adopted from the photon wave function approach. In the view of the result the optical soliton may be treated as a bound state (cluster) of many photons. \\   \\
PACS number(s): 42.65.k
\end{abstract}

\section{Introduction}
\label{sec: 1}

To describe photon in terms of a one-particle wave function, i.e., on the first quantization level, we follow the way presented in \cite{1, 2, 3}. Another approach was proposed in \cite{4, 4a}. The concept of the position-representation photon wave function has a long history and is still controversial. Nevertheless we do not want to discuss the question here. The reader interested in this problem is referred to \cite{1, 2, 3, 4, 4a, 4b, 4c} and references therein. In this paper the concept serves us as a convenient tool. 

The starting point for us is the free photon Schr{\"o}dinger equation \cite{1, 2, 3}. On this basis we develop the description of photon in medium in terms of the photon wave function. Some attempts of this kind have been presented in \cite{1, 2, 3}. What is new in our approach is to show that the influence of the medium on the photon can be described through some potentials. Though generally the idea is not new, see e.g. \cite{5, 6}, but here we realize it within the formalism of the photon wave function. This approach can be easily extended to the case of many interacting photons. For a large number of photons, it results as a nonlinear optics. In particular, taking as a simple model the photons interacting via Dirac delta-potential one obtains (in Hartree approximation) the Kerr type nonlinearity of the medium. The problem of interactions of  photons via delta-potential and the Kerr nonlinearity of the medium has its vast literature, especially in the context of quantization of nonlinear Schr{\"o}dinger equation (\cite{7, 8} and references therein). But all the literature refers to the procedure of creation-annihilation operators.   

The paper is organized in the following way. We begin the Section 2 with the brief description of a single photon in the medium. On this basis we extend the approach to the case of many interacting photons. We present that in Hartree approximation the many-photon quantum Schr{\"o}dinger equation takes on the form similar to the nonlinear Schr{\"o}dinger equation. The obtained equation expressed in terms of classical language is nothing else but  Maxwell equations with material relations in the form of Kerr type nonlinearity. In Section 3 we summarize results and give some final remarks.

\section{Results}
\label{sec: 2}

In the papers \cite{1, 2, 3} the following form of the Schr{\"o}dinger equation for free photon was proposed
\begin{equation}
\label{freeSchroed}
i \hbar \partial _{t} \mathcal{F} = H_{f} \mathcal{F},
\end{equation}

\begin{equation}
\label{freeFH}
\mathcal{F} =
\left[ \begin{array}{c}
\mathbf{E}(t,\mathbf{r}) + i  \mathbf{H}(t,\mathbf{r}) \\
\mathbf{E}(t,\mathbf{r}) - i  \mathbf{H}(t,\mathbf{r})
\end{array} \right] , \quad
H_{f} = c
\left[ \begin{array}{c c}
\mathbf{p} \cdot \mathbf{S}, & 0 \\
0, & -\mathbf{p} \cdot \mathbf{S}
\end{array} \right]   ,
\end{equation}

$\mathbf{p} = -i\hbar\nabla$ - momentum of photon;    $(\mathbf{S}_{i})_{kl}=-i\varepsilon_{ikl}$ - spin photon matrix.  
\newline
On the classical language, the equations are equivalent to the following Maxwell equations
\begin{equation}
\label{Max_rot}
\partial _{t} \mathbf{E} = c \nabla \times \mathbf{H}, \qquad
\partial _{t} \mathbf{H} = - c \nabla \times \mathbf{E}; \qquad \mathbf{D}=\mathbf{E}, \qquad \mathbf{B}=\mathbf{H}.
\end{equation}
describing free fields in vacuum. Since all the information carried by function $\mathcal{F}$  is contained in its positive energy (positive frequency) part $\mathcal{F}^{(+)}$, one may take this part as the true photon wave function \cite{3}
\begin{equation}
\label{psi}
\psi =\mathcal{F}^{(+)}.
\end{equation}
 To become a complete set of Maxwell equations, equation (\ref{Max_rot}) must be supplied by divergence conditions $\nabla\cdot\mathbf{E}=0$, $\nabla\cdot\mathbf{H}=0$. It is equivalent to the relation $\mathbf{p}\cdot\psi=0$. 

In order to describe the propagation of photon in dielectric, one should include in Hamiltonian the interaction term. On the microscopic level, such interaction is rather complicated, but here we will take it into account in a phenomenological way. We assume that the photon treated as a quantum object 'feels' the medium as an external classical field. For a stationary state the wave function takes on the form
\begin{equation}
\label{stacj}
\psi_{\omega} =\phi_{\omega}(\mathbf{r}) \exp(-i \omega t), \quad
\textrm{where} \quad \phi_{\omega}(\mathbf{r})=
\left[ \begin{array}{c}
\mathbf{E}(\mathbf{r}) + i  \mathbf{H}(\mathbf{r}) \\
\mathbf{E}(\mathbf{r}) - i  \mathbf{H}(\mathbf{r})
\end{array} \right].
\end{equation}
Generally, the couplings of the medium with the electric and magnetic part of the wave function may be different. To take it into account we introduce two real and symmetric matrices $\gamma$ and $\eta$ which split the wave function $\psi_{\omega}$ into electric and magnetic parts
\begin{equation}
\label{rozszcz}
\gamma \psi_{\omega} + \eta \psi_{\omega} =\psi_{\omega},
\end{equation}
\begin{equation}
\label{rozszczEH}
\gamma \psi_{\omega}=
\left[ \begin{array}{c}
\mathbf{E} \\
\mathbf{E}
\end{array} \right],   \qquad
\eta \psi_{\omega} =
\left[ \begin{array}{c}
i\mathbf{H} \\
-i\mathbf{H}
\end{array} \right],
\end{equation}
\begin{equation}
\label{gameta}
\gamma  = \frac{1}{2}
\left[ \begin{array}{c c}
1 & 1 \\
1 & 1
\end{array} \right], \qquad
\eta = \frac{1}{2}
\left[ \begin{array}{c c}
1 & -1 \\
-1 & 1
\end{array} \right].
\end{equation}
The projection operators $\gamma$, $\eta$ fulfill the following relations i.e.: $\gamma^{2}=\gamma, {~}\eta^{2}=\eta, {~} \gamma \eta =0$. 
When the couplings with the electrical and magnetic parts are taken into account, the Schr{\"o}dinger equation takes on the form
\begin{equation}
\label{dieom}
\hbar \omega \psi_{\omega} = H_{f} \psi_{\omega} - \Omega_{\omega}
(\mathbf{r}) \gamma \psi_{\omega}.
\end{equation}
$\Omega_{\omega}(\mathbf{r}) \gamma$, $\Gamma_{\omega}(\mathbf{r}) \eta$  have interpretation of potential energy operators. $\Omega_{\omega}(\mathbf{r})$ is connected with the dielectric susceptibility $\chi_{\omega}(\mathbf{r})=\Omega_{\omega}(\mathbf{r})/\hbar\omega$, and $\Gamma_{\omega}(\mathbf{r})$ with magnetic susceptibility $\chi^{m}_{\omega}(\mathbf{r})=\Gamma_{\omega}(\mathbf{r})/\hbar\omega$.

And now consider $N$ interacting photons propagating in a homogeneous medium. Let the Hamiltonian $H$ has the form 
\begin{equation}
\label{hamiltonian fotonow}
H=\sum_{l=1}^{N} H_{l}-N\Omega_{\omega }\gamma - \sum_{j>l}H_{lj}.
\end{equation}
where the first term is the sum of $N$ identical one-particle Hamiltonians  in the form given by equation (\ref{freeFH}) - each describing evolution of a free photon, the second term describes their coupling to the non-magnetic medium ($\Gamma_{\omega}=0$),  and the third term is the sum taken over all the  photon-photon interactions $H_{lj}$. It may be any mechanism leading effectively to the weak interaction between photons. The main assumption is that the interaction can be modeled by Dirac $\delta$-function (when the two photons are simultaneously in a very small volume element $V$ of the space). The measure of its intensity is given by a parameter $A_{\omega}$. Because the photon-photon interaction is performed with the help of the medium, and the dielectric medium acts effectively only on the electrical component of the wave function, therefore the matrix $\gamma$ should appear in the Hamiltonian $H_{lj}$:
\begin{equation}
\label{czlon oddzialywania}
H_{lj}=A_{\omega} V \gamma \delta(\mathbf{r}_{l}-\mathbf{r}_{j}).
\end{equation}
 
The Schr{\"o}dinger equation for the $N$ photons takes on the form
\begin{equation}
\label{rownanie Schrodingera N fotonow}
i\hbar\partial_{t}\psi(\mathbf{r}_{1},\ldots,\mathbf{r}_{N},t)=H\psi
(\mathbf{r}_{1},\ldots,\mathbf{r}_{N},t).
\end{equation}
The position-representation $N$-photon wave function $\psi(\mathbf{r}_{1},\ldots,\mathbf{r}_{N},t)$ is constructed in a similar way as the one constructed previously for one photon
\begin{equation}
\label{funkcja falowa N fotonow}
\psi=\left[ \begin{array}{c}
\mathcal{E}(t,\mathbf{r}_{1},\ldots,\mathbf{r}_{N})+i\mathcal{H}(t,\mathbf{r}_{1},\ldots,\mathbf{r}_{N})
\\
\mathcal{E}(t,\mathbf{r}_{1},\ldots,\mathbf{r}_{N})-i\mathcal{H}(t,\mathbf{r}_{1},\ldots,\mathbf{r}_{N})
\end{array}\right] =\phi_{\omega}(\mathbf{r}_{1},\ldots,\mathbf{r}_{N})\exp(-i\omega t).
\end{equation}
The last equality means that only stationary states are considered.  In Hartree approximation, see e.g. \cite{9}, $\phi_{\omega}(\mathbf{r}_{1},\ldots ,
\mathbf{r}_{N})$ can be written as a product of one-photon wave functions $\phi_{\omega}(\mathbf{r}_{k})$  ($k=1,\ldots,N$)
\begin{equation}
\label{rozklad funkcji falowej na funkcje jednofotonowe}
\phi_{\omega}(\mathbf{r}_{1},\ldots,\mathbf{r}_{N}){~}={~}\prod_{i=1}^{N}\phi_{\omega}(\mathbf{r}_{i}){~}=
\end{equation}
\begin{displaymath}
=\left[ \begin{array}{c}
\mathbf{e}(\mathbf{r}_{1})+i\mathbf{h}(\mathbf{r}_{1}) \\
\mathbf{e}(\mathbf{r}_{1})-i\mathbf{h}(\mathbf{r}_{1})
\end{array} \right] \quad \ldots \quad
\left[ \begin{array}{c}
\mathbf{e}(\mathbf{r}_{N})+i\mathbf{h}(\mathbf{r}_{N}) \\
\mathbf{e}(\mathbf{r}_{N})-i\mathbf{h}(\mathbf{r}_{N})
\end{array} \right] \equiv
\end{displaymath}
\begin{displaymath}
{~}\left[ \begin{array}{c}
[e_{x}(\mathbf{r}_{1})+ih_{x}(\mathbf{r}_{1})]\cdots[e_{x}(\mathbf{r}_{N})+ih_{x}(\mathbf{r}_{N})]
\\{}
[e_{y}(\mathbf{r}_{1})+ih_{y}(\mathbf{r}_{1})]\cdots[e_{y}(\mathbf{r}_{N})+ih_{y}(\mathbf{r}_{N})]
\\{}
[e_{z}(\mathbf{r}_{1})+ih_{z}(\mathbf{r}_{1})]\cdots[e_{z}(\mathbf{r}_{N})+ih_{z}(\mathbf{r}_{N})] \\{}
[e_{x}(\mathbf{r}_{1})-ih_{x}(\mathbf{r}_{1})]\cdots[e_{x}(\mathbf{r}_{N})-ih_{x}(\mathbf{r}_{N})]
\\{}
[e_{y}(\mathbf{r}_{1})-ih_{y}(\mathbf{r}_{1})]\cdots[e_{y}(\mathbf{r}_{N})-ih_{y}(\mathbf{r}_{N})]
\\{}
[e_{z}(\mathbf{r}_{1})-ih_{z}(\mathbf{r}_{1})]\cdots[e_{z}(\mathbf{r}_{N})-ih_{z}(\mathbf{r}_{N})]
\end{array} \right].
\end{displaymath}

For many photons, the three-dimensional vectors $\mathcal{E}$, $\mathcal{H}$   by no means can be treated as a classical electric and magnetic field (only in the special case of one photon the vectors  have this interpretation) . In general, our point of view is that in many photon case, the vectors of an effective, self-consistent one-photon wave function could be interpreted as a classical electric and magnetic field.

Multiplying (\ref{rownanie Schrodingera N fotonow}) by the product 
\begin{equation}
\label{iloczyn jednofotonowych funkcji falowych}
\prod_{i=2}^{N}\phi_{\omega}(\mathbf{r}_{i}),
\end{equation} 
taking integral over coordinates of the particles over all the space 
\begin{equation}
\label{calkowanie po wspolrzednych}
\int d\mathbf{r}_{2} \ldots d\mathbf{r}_{N}
\end{equation}
and  normalizing  the one-particle function
\begin{equation}
\label{normalizacja jednofotonowej funkcji falowej}
\int \phi_{\omega}^{' \dagger}(\mathbf{r}) \phi_{\omega}^{'}(\mathbf{r})
d\mathbf{r}=1, \quad \textrm{where}\quad
\phi_{\omega}^{'}(\mathbf{r})=s \phi_{\omega}(\mathbf{r})
\end{equation}
($s$ is a normalizing factor),
one obtains the one-particle self-consistent equation
\begin{equation}
\label{samouzgodnione rownanie falowe}
\hbar\omega\phi=H_{f}\phi-\Omega^{\ast}_{\omega}\gamma\phi-(N-1)A_{\omega}Vs^{2}(\phi^{\dagger}\phi)\gamma\phi.
\end{equation}
$\phi^{\dagger}$ means Hermitian adjoint of $\phi$.
$H_{f}$ is one-particle Hamiltonian given by (\ref{freeFH}).
$\Omega^{\ast}_{\omega}$ is a normalized coupling
\begin{equation}
\label{Omega z asteriksem}
\Omega^{\ast}_{\omega}=\Omega_{\omega}+2 (N-1) A_{\omega} V s^{2} \int
[\mathbf{e}^{2}(\mathbf{r})+\mathbf{h}^{2}(\mathbf{r})] \mathbf{e}^{2}
(\mathbf{r}) d \mathbf{r}.
\end{equation}
Note that in equation (\ref{samouzgodnione rownanie falowe}) the unnormalized wave function $\phi$ appears. The unnormalized wave function is more convenient for interpretation in terms of classical electric and magnetic fields. To this end one may recast equation (\ref{samouzgodnione rownanie falowe}) to ordinary form of Maxwell equations 
\begin{equation}
\label{samouzgodnione jako rownania Maxwella}
\frac{\omega}{c}\, [\varepsilon^{\ast}_{\omega} + \alpha_{\omega}(N-1)
(\mathbf{e}^{2}+\mathbf{h}^{2})]\mathbf{e}=i\nabla\times\mathbf{h},
\qquad \frac{\omega}{c}\mathbf{h}=-i\nabla\times\mathbf{e},
\end{equation}
\begin{displaymath}
\textrm{where}\quad \varepsilon^{\ast}_{\omega}=1+\frac{\Omega^{\ast}_
{\omega}}{\hbar\omega}, \qquad
\alpha_{\omega}=\frac{2 A_{\omega} V s^{2}}{\hbar\omega}.
\end{displaymath}
The quantities $\mathbf{e}$ and $\mathbf{h}$ require some renormalization
\begin{equation}
\label{renormalizacja pol e, h}
\mathbf{E}_{kl}=\sqrt{N-1}{~}\mathbf{e}, \qquad \mathbf{H}_{kl}=\sqrt{N-1}{~}\mathbf{h}.
\end{equation}
Thus the equation (\ref{samouzgodnione jako rownania Maxwella}) takes on the form
\begin{equation}
\label{samouzgodnione jako rownania Maxwella - klasyczne}
\frac{\omega}{c}[\varepsilon_{\omega} + \alpha_{\omega}(\mathbf{E}_{kl}
^{2}+\mathbf{H}_{kl}^{2})]\mathbf{E}_{kl}=i\nabla\times\mathbf{H}_{kl},
\qquad \frac{\omega}{c}\mathbf{H}_{kl}=-i\nabla\times\mathbf{E}_{kl},
\end{equation}
Note that  in this equation  
$\varepsilon^{\ast}_{\omega}$ is replaced by the quantity $\varepsilon_{\omega}$ defined as
\begin{equation}
\varepsilon_{\omega}=1+\frac{\Omega_{\omega}}{\hbar\omega}.
\end{equation}
It is due to the fact that in definition (\ref{Omega z asteriksem}) of 
$\Omega^{\ast}_{\omega}$  in the second term the factor  $N-1$ appears and fields in integral are in fourth power.
Therefore after the renormalization  (\ref{renormalizacja pol e, h}) the fields in the integral are divided by $N-1$. The procedure of the mean field approximation (Hartree) is correct only for large $N$. It means that the second term in  (\ref{Omega z asteriksem}) can be omitted and one thus obtains
\begin{equation}
\label{rownosc asteriksow  bez asteriksow}
\Omega^{\ast}_{\omega}=\Omega_{\omega}, \qquad \textrm{and} \qquad
\varepsilon^{\ast}_{\omega}=\varepsilon_{\omega}.
\end{equation}

For electromagnetic field (in medium as well as in vacuum) one may write 
\begin{equation}
\label{proporcjonalnosc E^2 do H^2}
\mathbf{E}_{kl}^2 \propto \mathbf{H}_{kl}^2.
\end{equation}
Therefore
\begin{equation}
\label{rownania Maxwella - klasyczne}
\frac{\omega}{c}\varepsilon_{NL}\mathbf{E}_{kl}=i\nabla\times
\mathbf{H}_{kl}, \qquad \frac{\omega}{c}\mathbf{H}_{kl}=-i\nabla\times
\mathbf{E}_{kl},
\end{equation}
\begin{equation}
\label{niel}
\textrm{where} \qquad \varepsilon_{NL}=\varepsilon_{\omega}+
r_{\omega}\alpha_{\omega}\mathbf{E}_{kl}^2.
\end{equation}
$r_{\omega}$ is a factor refering to the relation
(\ref{proporcjonalnosc E^2 do H^2}).

The interaction between photons results as the Kerr type nonlinearity of the medium (\ref{niel}). Such type of nonlinearity is required to obtain (in paraxial and slowly varying envelop approximation) soliton solutions of Maxwell equations \cite{10}. In the view of the above results the soliton can be treated as a bound state (cluster) of many photons. This is the attracting interaction between photons, which can counteract spreading (because of dispersion or diffraction) of the wave packet.

\section{Summary}
\label{sec: 3}

The influence of the medium on photon can be described by some scalar potentials. This approach can be easily extended to the case of many interacting photons. Taking the simplest model of such interaction (via delta-potential) in Hartree approximation one obtains Kerr type nonlinearity of the medium. One may say that optically nonlinear material it is the material that enables interaction between photons. This result suggests that optical solitons  are nothing else but clusters of interacting photons. In such a cluster the attraction between photons counteracts dispersion or diffraction. This is a similar point of view to that presented in \cite{11} where solitons of magnetization in magnetics were described as clusters of interacting (via delta-potential) magnons.

\end{document}